\documentclass[10pt,twocolumn,letterpaper]{article}

\usepackage{wacv}              
\usepackage{cuted}

\usepackage{multirow}
\usepackage{makecell}
\usepackage{booktabs}

\usepackage{times}
\usepackage{latexsym}
\usepackage{booktabs}
\usepackage[table]{xcolor}
\usepackage{makecell}
\usepackage{amsmath}
\usepackage{pifont}
\usepackage{longtable}
\usepackage{array}
\usepackage{multicol}
\usepackage{ragged2e}
\usepackage[T1]{fontenc}
\usepackage{xcolor}

\usepackage[utf8]{inputenc}
\usepackage{textcomp}
\usepackage{amssymb}
\usepackage{microtype}

\usepackage{xr-hyper}
\newcommand{\beginsupplement}{%
  \clearpage
  \setcounter{page}{1}\renewcommand{\thepage}{S\arabic{page}}%
  \setcounter{section}{0}\renewcommand{\thesection}{S\arabic{section}}%
  \setcounter{figure}{0}\renewcommand{\thefigure}{S\arabic{figure}}%
  \setcounter{table}{0}\renewcommand{\thetable}{S\arabic{table}}%
  \setcounter{equation}{0}\renewcommand{\theequation}{S\arabic{equation}}%
}
\definecolor{wacvblue}{rgb}{0.21,0.49,0.74}
\usepackage[pagebackref,breaklinks,colorlinks,allcolors=wacvblue]{hyperref}

\def\wacvPaperID{1024} 
\def\confName{WACV}
\def\confYear{2027}

\title{WebAgentGuard: A Multimodal Reasoning-Driven Guard Model for Detecting Prompt Injection Attacks in Web Agents}

\author{
Yulin Chen$^{1}$\thanks{Yulin Chen and Tri Cao contributed equally.},
Tri Cao$^{1}$\footnotemark[1],
Haoran Li$^{2}$,
Yibo Li$^{1}$,
Yue Liu$^{1}$,\\
Yufei He$^{1}$,
Khoi M. Le$^{1}$,
Yangqiu Song$^{2}$,
Shuicheng Yan$^{1}$,
Bryan Hooi$^{1}$\\[6pt]
$^{1}$National University of Singapore \\
$^{2}$Hong Kong University of Science and Technology \\
\texttt{\{chenyulin28,caotri\}@u.nus.edu}\\
}

\begin{document}
\maketitle
\begin{abstract}
Web agents powered by vision–language models (VLMs) enable autonomous interaction with web environments by perceiving and acting on both visual and textual webpage content to accomplish user-specified tasks. However, they are highly vulnerable to prompt injection attacks, where adversarial instructions embedded in HTML or rendered screenshots can manipulate agent behavior and lead to harmful outcomes such as information leakage. Existing defenses, including system prompt defenses and direct fine-tuning of agents, have shown limited effectiveness. To address this issue, we propose a defense framework in which a web agent operates in parallel with a dedicated guard agent, decoupling prompt injection detection from the agent’s own reasoning. Building on this framework, we introduce \textbf{WebAgentGuard}, a reasoning-driven, multimodal guard model for prompt injection detection. We construct a synthetic multimodal dataset using GPT-5 spanning 164 topics and 230 visual and UI design styles, and train the model via reasoning-intensive supervised fine-tuning followed by reinforcement learning. Experiments across multiple benchmarks show that WebAgentGuard consistently outperforms strong baselines while preserving agent utility, without introducing additional latency.\footnote{The code is available at \url{https://github.com/LukeChen-go/webagentguard}}
\end{abstract}


\begin{figure*}[t]
    \centering
    \includegraphics[width=0.9\linewidth]{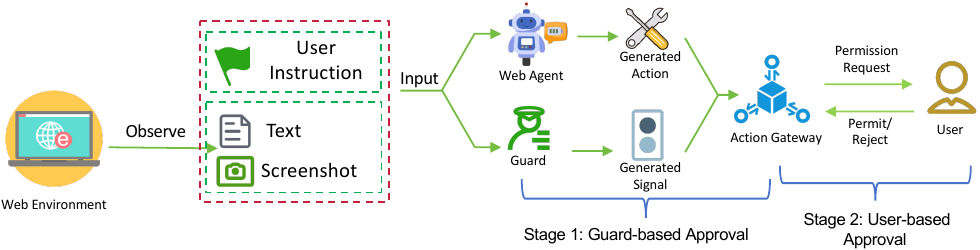}
    \caption{Overview of the parallel guard framework. In each execution loop, the agent observes the webpage (screenshot and text), which are sent to both the web agent and the guard along with the user instruction. The web agent proposes the next action, while the guard outputs a permission signal. At the Action Gateway, the proposed action is executed if approved by the guard (Stage 1). If the guard denies permission, the system prompts the user for manual approval; if approved, the action proceeds, otherwise the execution trajectory is terminated (Stage 2).}
    \vspace{-5pt}
    \label{fig:intro}
\end{figure*}
\section{Introduction}

Web agents are AI systems designed to automate time-consuming online tasks by interacting with web environments, leveraging vision–language models (VLMs) to interpret both visual and textual webpage content \cite{ning2025survey, zhou2023webarena, wei2025webagent, wu2025webdancer}. Despite their utility, web agents are highly vulnerable to prompt injection attacks, where adversarial instructions embedded in webpage elements such as HTML content or rendered screenshots can manipulate agent behavior and lead to harmful outcomes, including information leakage \cite{zhang2025attacking, evtimov2025wasp, cao2025vpi, liao2025eia}. Prior defenses, including fine-tuning VLM agents and system-prompt defenses, have proven largely ineffective, with attack success rates exceeding 80\% in many cases \cite{cao2025vpi}. A key reason is that the agent’s reasoning is tightly coupled with task completion, prioritizing instruction following over safety analysis and making prompt injection difficult to detect during execution. 


To address this issue, we consider decoupling injection detection from the agent’s reasoning by proposing a defense framework, shown in Figure~\ref{fig:intro}, where a web agent runs in parallel with a dedicated guard agent. In practice, guard models are much smaller than web agents, allowing the verification signal to be produced before the agent finishes reasoning and issues the next action, thereby preserving overall execution efficiency. We note that several guard models have been proposed~\cite{chi2024llamaguard3vision, zhao2025qwen3guard, liu2025guardreasoner, liu2025guardreasoner-vl, li2025gspr, zheng2025rsafe}; however, they are primarily trained and evaluated for jailbreak attacks rather than web-based prompt injection scenarios.

Building on the parallel guard framework, we propose WebAgentGuard, a multi-modal, reasoning-driven prompt injection guard specifically designed for web agents. We construct a synthetic multimodal dataset using GPT-5 that jointly captures HTML content, rendered visual context, and user intent. The dataset spans 164 topics and 230 visual and UI design styles, and includes paired benign and injected webpages created by injecting adversarial instructions into HTML while preserving the original user instruction. To explicitly train reasoning capability, each fine-tuning sample is augmented with step-by-step reasoning traces generated by GPT-5 and corresponding ground-truth labels. We first cold-start the guard model via supervised fine-tuning on this reasoning-annotated data, and then further refine it using Group Relative Policy Optimization (GRPO) \cite{shao2024deepseekmath}.

We conduct comprehensive experiments to evaluate the performance of WebAgentGuard. On our crafted evaluation dataset, WebAgentGuard achieves nearly 100\% recall, substantially outperforming baseline methods, which exhibit poor detection performance. For out-of-domain evaluation, we benchmark against VPI-Bench~\cite{cao2025vpi} and EIA~\cite{liao2025eia}, and reproduce the PopUp Attack~\cite{zhang2025attacking} on top-ranked websites collected from SimilarWeb \footnote{https://www.similarweb.com/top-websites/}. Across these benchmarks, WebAgentGuard consistently outperforms baselines, achieving over 90\% average recall. In utility evaluation, WebAgentGuard preserves strong performance on the WebArena ~\cite{zhou2023webarena}. Finally, we show that WebAgentGuard introduces lower latency than the web agent at each reasoning step and does not negatively affect overall execution efficiency. Our contributions are summarized as follows:
\begin{itemize}
\item We propose a parallel defense framework in which a guard model runs alongside the web agent, decoupling safety reasoning from task execution for prompt injection detection.
\item We construct a synthetic multimodal training dataset for prompt injection defense spanning diverse web topics and styles.
\item We train WebAgentGuard, a reasoning-driven guard model that effectively detects prompt injection attacks, consistently outperforming strong baselines while preserving web agent utility and execution efficiency.
\end{itemize}
\section{Related Work}
\label{sec:related}

\paragraph{Prompt Injection Attacks.}
Prompt injection attack occurs when an agent unintentionally processes adversarial instructions embedded within external content encountered during task execution, without malicious intent from the user \citep{greshake2023more, debenedetti2024agentdojo, fu2024imprompter, wang2025manipulating, chen2025topicattack, chen2025backdoor}. These attacks exploit the agent’s implicit trust in environmental observations, including webpages, images, and tool outputs, to covertly influence its reasoning and subsequent actions. Prior work has demonstrated the effectiveness of prompt injection against web agents via multiple attack strategies, such as injected HTML content \citep{wu2024wipi, cao2025phishagent, li2024knowphish}, adversarial images \citep{wu2024adversarial, fu2024imprompter, aichberger2025attacking}, malicious webpages that induce private data leakage \citep{xu2024advweb, liao2025eia}, and dynamic tool-mediated interactions in interactive environments like AgentDojo \citep{debenedetti2024agentdojo}. More recent studies begin to explore visually grounded attack methods, including pop-up-based manipulations \citep{zhang2025attacking} and pixel-level perturbations introduced during webpage rendering, as proposed in WebInject \citep{wang2025webinject}. Notably, VPI-Bench \cite{cao2025vpi} further shows that state-of-the-art web agents can be manipulated into executing complex, multi-step malicious behaviors, explicitly targeting both user security and system-level vulnerabilities.

\paragraph{Defense Methods.}
Numerous defenses have been proposed to mitigate \emph{prompt injection attacks}, including aligning the model with instruction hierarchy~\citep{wallace2024instruction, chen2025struq}.  These approaches enforce the model to follow higher-level instructions when there are instruction conflicts.
However, defending against \emph{prompt injection attacks on agents} remains challenging: recent efforts fine-tune LLM/VLM agents against prompt injection~\citep{anthropic2024} and deploy system prompt defenses, yet these approaches remain ineffective against  prompt injection~\citep{cao2025vpi} on agents, as agents continue to exhibit high attack success rates. More recently, researchers introduce guard models as runtime defenses that primarily target \emph{jailbreak attacks}~\citep{OpenAIModeration,Llamaguard,liu2025guardreasoner,wang2024adashield,sun2024safeguarding,zhang2023jailguard,oh2024uniguard,Vlmguard, wu2025automating}. However, these guard models do not effectively protect against indirect prompt injection. In this work, we propose a guard model for web agents that explicitly mitigates such vulnerabilities.
Compared with prior guard-agent and safety-shielding work, our work differs across three fundamental dimensions. First, it introduces a parallel reasoning-driven operational paradigm that runs concurrently with the agent to detect injections in real time, avoiding the latency of intermediate-reasoning interventions like MirrorAgent \cite{guo2025mirror}. Second, it expands the protection scope specifically to dynamic, multimodal web-agent environments, overcoming the textual policy constraints of GuardAgent \cite{xiang2024guardagent}, the text-only limitations of BrowseSafe \cite{zhang2025browsesafe}, and the tool-centric focus of VIGIL \cite{lin2026vigil}. Third, it maximizes deployment efficiency through unified reasoning-based detection, eliminating the heavy computational overhead and repeated model calls required by segment-level approaches like WebSentinel \cite{wang2026websentinel}.
\begin{figure*}
    \centering
    \includegraphics[width=0.9\linewidth]{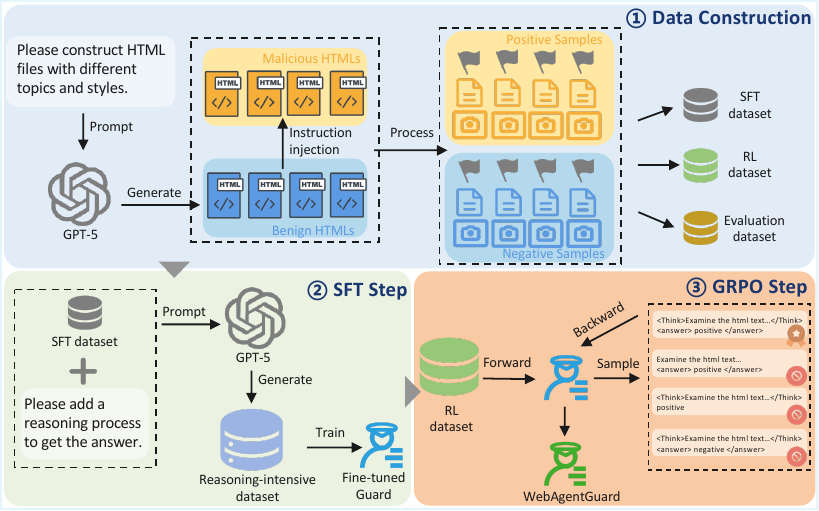}
    \caption{Overview of the training pipeline.
\ding{172} We prompt the GPT model to generate HTML files and construct positive and negative samples containing user instructions, screenshots, and processed HTML files, which are subsequently divided into SFT, RL, and evaluation datasets.
\ding{173} GPT-5 is further used to inject reasoning chains into the SFT data to cold-start the backbone VLM.
\ding{174}  The model is then post-trained with GRPO on the RL dataset.}
    \label{fig:method}
    \vspace{-10pt}
\end{figure*}

\section{Problem Formulation}
\paragraph{Web Agent Working Pipeline.}
For a web agent, a user provides an instruction or request \( I \), after which the agent interacts with the web environment. At step \( t \), the agent receives an observation \( ob_t \) of the environment \( e_t \), defined as
$ob_t = \mathcal{O}(e_t).$
The observation \( ob_t \) consists of a webpage screenshot \( S_t \) and a processed HTML representation \( T_t \), i.e.,
$ob_t = \{ S_t, T_t \}$.
Based on the user instruction \( I \), the observation \( ob_t \), and the internal state from the previous step \( s_{t-1} \), which contains the agent’s memory and interaction history, the agent updates its internal state as
$s_t = \mathcal{U}(I, s_{t-1}, ob_t)$.
The agent then generates the next action according to its policy
$a_t = \pi(s_t)$,
and executes the action to transition to a new environment:
$e_{t+1} = \mathcal{T}(e_t, a_t)$.

\paragraph{Prompt Injection Attacks.}
In a prompt injection attack, an attacker injects malicious instructions into the raw HTML content of a webpage, ensuring that the agent can perceive the injected instructions through the textual modality \( T^m \), the visual modality \( S^m \), or both. Suppose that at step \( t \), the agent visits a malicious website crafted by the attacker and receives a malicious observation
$ob_t^{m} = \mathcal{O}(e_t^{m})$.
This observation drives the agent into a malicious internal state:
$s_t^{m} = \mathcal{U}(I, s_{t-1}, ob_t^{m})$,
causing the agent to follow the injected instructions and generate a harmful action
$a_t^{m} = \pi(s_t^{m})$,
such as leaking the user’s API key to the attacker. After executing the action,
the agent ultimately causes damage to the user.

\section{Methodology}

\subsection{Defense with WebAgentGuard.}
To defend against prompt injection attacks, we train and leverage a guard model, {WebAgentGuard}, denoted as \( \mathcal{G} \). As shown in Figure \ref{fig:intro}, at each step, the guard model analyzes the agent’s observation (including screenshots and HTMLs) and produces a binary decision signal:
$g_t = \mathcal{G}(ob_t)$,
which determines whether the agent is permitted to execute the proposed action. The guard operates in parallel with the agent’s reasoning process \( \pi \). An action is executed only if the guard approves it (\( g_t = 1 \)), or if the guard rejects it (\( g_t = 0 \)) but explicit human authorization is provided (\( h_t = 1 \)). Formally, the environment transition is defined as
\begin{equation}
\small
    e_{t+1} =
\begin{cases}
\mathcal{T}(e_t, a_t),
& \text{if } g_t = 1 \lor (g_t = 0 \land h_t = 1), \\
\text{END},
& \text{otherwise}.
\end{cases}
\end{equation}

``\text{END}'' means the termination of the task. This design decouples safety verification from task-oriented reasoning, preventing task completion from influencing prompt injection detection.
To build WebAgentGuard, we first construct the training data and then train the backbone VLM on the synthesized data. The overall pipeline is illustrated in Figure~\ref{fig:method}.

\subsection{Data Curation}

\paragraph{Sample Synthesis.}
Training a reliable guard model for web agents requires data that captures web content with diverse topics and design styles in multimodal settings. However, collecting such data at scale from real-world websites is challenging due to limited access and limited controllability. To address these challenges, we adopt a fully synthetic yet structured data synthesis pipeline that enables both high diversity and realistic content generation.
We synthesize the web content by leveraging GPT-5.
To ensure diversity and coverage, we prompt GPT-5 to generate webpages spanning different topics
and website design styles.
Specifically, we use the following prompt:

\begin{quote}
\small
\textit{``Help me design an HTML website about \texttt{<topic>}, with the style of \texttt{<style>} in English.
Please ensure that all images in the HTML file are valid and visible.
Provide only the HTML code.''}
\end{quote}

Here, \texttt{<topic>} is instantiated with various topics collected from SimilarWeb, covering 24 topic categories and 164 topics,  as shown in Table \ref{tab:topics} in the supplementary material,
and \texttt{<style>} is replaced with diverse website design styles spanning 11 visual style categories and 230 fine-grained design styles, generated by GPT-5 and detailed in Table \ref{tab:styles} in the supplementary material.
For each topic, we generate 20 HTML files with randomized styles. We then randomly sample 5 files per topic to verify that they are benign and safe. Once validated, these generated files are collected as negative samples.
We capture the corresponding screenshot for each HTML file.
Since CSS code, JavaScript descriptions, and HTML tag attributes are not directly perceived by web agents
but significantly increase input length, we remove these redundant components and retain only the
essential content that is visually displayed, such as textual elements.
We then generate a user instruction related to the webpage content using GPT-5 {with the prompt shown in Figure \ref{fig:prompt_user} in the supplementary material}.
As a result, each negative sample consists of four elements:
a user instruction $I$, a processed benign HTML file $T^b$,
a benign screenshot $S^b$, and a negative label $l^b$.
The negative dataset is defined as $D^{\text{neg}} = \{ I_i, T^b_i, S^b_i, l^b \}_{i=1}^{N}.$

To construct positive samples, we inject instructions from ~\citet{alpaca}
into the benign HTML files generated by GPT-5 at random positions.
This strategy allows us to simulate realistic prompt injection attacks that are embedded within webpage content.
Following the same pipeline as in negative sample construction,
we collect screenshots in which the injected instructions are visually present.
These HTML files are further post-processed to preserve only the key components observable by the agent.
Each positive sample therefore includes a user instruction $I$,
a processed malicious HTML file $T^m$,
a malicious screenshot $S^m$, and a positive label $l^m$,
forming the positive dataset
$D^{\text{pos}} = \{ I_i, T^m_i, S^m_i, l^m \}_{i=1}^{N}.$

Finally, we combine the negative and positive datasets and divide them into
the supervised fine-tuning dataset $D^{\text{SFT}}$,
the reinforcement learning dataset $D^{\text{RL}}$,
and the evaluation dataset $D^{\text{eval}}$.

\paragraph{Reasoning Construction.}
For the SFT dataset $D^{\text{SFT}}$, our goal is to cold-start the VLM by teaching it how to reason before making a final decision and produce outputs using a structured template,
where the reasoning process is enclosed within ``\texttt{<think>...</think>}'' tags, and the final decision is enclosed within ``\texttt{<answer>...</answer>}'' tags.
To construct such reasoning-intensive training data, we prompt GPT-5 to generate the reasoning and come to the ground truth answer based on the reasoning process using the format.
The prompt used is provided in Figure \ref{fig:prompt_reasoning}  in the supplementary material. The reasoning steps enable the model to jointly consider the user instruction, the rendered webpage screenshot, and the processed HTML content, and to explicitly articulate its reasoning process before producing a final binary judgment. After generating the reasoning-intensive data, we filter out samples that either leak the ground-truth answer during the reasoning process or produce incorrect final answers.
This results in a cleaned SFT dataset with reasoning, denoted as $D^{\text{SFT-R}} = \{I_i, T_i, S_i, y_i\}_{i=1}^{N_{\text{sft}}}$,
where $y_i$ contains the templated reasoning and answer generated by GPT-5, and $N_{\text{sft}}$ is the total number of training samples.
Statistics of the resulting dataset splits are shown in {Table} \ref{tab:dataset_stats}.

\begin{table}[h]
\centering
  \scriptsize
  \setlength{\tabcolsep}{15pt}
\begin{tabular}{lccc}
\hline
\textbf{Dataset} & \textbf{Positive} & \textbf{Negative} & \textbf{Total} \\ \hline
SFT        & 938  & 983  & 1,921 \\
RL         & 1,675 & 1,779 & 3,454 \\
Evaluation & 500  & 500  & 1,000 \\
\hline
\end{tabular}
\caption{Statistics of SFT, RL and evaluation datasets.}
\vspace{-10pt}
\label{tab:dataset_stats}
\end{table}



\subsection{Two-stage Training Framework}
\paragraph{Cold-Start SFT.}
To help the backbone VLM learn how to reason for predictions and generate outputs in the required template format,
we first fine-tune it using the constructed reasoning dataset $D^{\text{SFT-R}}$.
The input to the model consists of a tuple: user instruction $I$, processed HTML file $T$, and screenshot $S$;
the target output is the templated reasoning sequence $y$ generated by GPT-5.
The model is trained by minimizing the standard language modeling loss, as shown in the equation below:
\begin{equation}
\small
\mathcal{L}
= - \sum_{i=1}^{N_{\mathrm{sft}}}
\log P_\theta \!\left(
y_i \,\middle|\,
I_i,\, T_i,\, S_i\
\right)
\end{equation}

\paragraph{Post-training with RL.}
Supervised fine-tuning (SFT) provides a strong initialization for adapting the backbone VLM to the target task. However, this training step is inherently bounded by the training data, leaving the model unable to make robust decisions in unseen scenarios. To overcome this limitation, we further align the model using reinforcement learning (RL) to encourage exploration and learn effective decision-making strategies in unfamiliar environments.
We adopt Group Relative Policy Optimization (GRPO), which improves policy performance by comparing relative outcomes among multiple sampled trajectories. The GRPO objective is defined as:
\begin{equation}
\small
\begin{aligned}
\mathcal{J}_{\text{GRPO}}(\theta) = 
&\ \mathbb{E}_{q \sim P(Q), \{o_i\}_{i=1}^{|G|} \sim \pi_{\theta_{\text{old}}}(O|q)} \Bigg[ \\
&\quad \frac{1}{|G|} \sum_{i=1}^{|G|} \Bigg( \min \left( \frac{\pi_\theta(o_i|q)}{\pi_{\theta_{\text{old}}}(o_i|q)} A_i, \right. \\
&\qquad\quad \left. \text{clip} \left( \frac{\pi_\theta(o_i|q)}{\pi_{\theta_{\text{old}}}(o_i|q)}, 1 - \epsilon, 1 + \epsilon \right) A_i \right) \\
&\quad - \beta \mathbb{D}_{\text{KL}}(\pi_\theta \| \pi_{\text{ref}}) \Bigg]
\end{aligned}
\end{equation}
Here, $q$ denotes the input prompt (identical to SFT), including the user instruction, processed HTML, and screenshot. $G$ represents a group of outputs sampled from the old policy $\pi_{\theta_{\text{old}}}$, and $\pi_{\text{ref}}$ is the SFT-trained VLM. The KL term regularizes policy updates, with $\beta$ controlling its strength.

Each sampled output corresponds to a complete reasoning process followed by a final prediction. The advantage $A$ is computed via group-wise normalization:
\begin{equation}
\small
A_i = \frac{r_i - \text{mean}(\{r_1, r_2, \cdots, r_G\})}{\text{std}(\{r_1, r_2, \cdots, r_G\})}
\end{equation}

Here, $r$ denotes the reward assigned to the model's output $o$.  
Regarding reward design, because our reinforcement learning objective is to improve the model's decision-making in unseen scenarios, we focus primarily on the final prediction. Therefore, we design a simple, rule-based reward function as follows:
a reward of 1 is given if and only if the VLM-generated output adheres to the required reasoning and answering format. Specifically, if it includes properly filled spans of 
``\texttt{<think>...</think>}'' and ``\texttt{<answer>...</answer>}'' and the final answer matches the ground-truth label, the reward is 1.
Otherwise, a reward of 0 is assigned.

\section{Experiments}

\label{sec:exp}
\subsection{Experimental Settings}


\paragraph{Training Configurations.}
For the SFT stage, we implement our code based on LlamaFactory~\cite{zheng2024llamafactory} using two NVIDIA H200 GPUs.
We set the total batch size to 16, with a batch size per GPU of 8, and a maximum sequence length of 8192.
The model is trained for 3 epochs using the Adam optimizer with a learning rate of $5 \times 10^{-6}$.

For the GRPO reinforcement learning stage, we build our implementation on the VERL package~\citep{sheng2025hybridflow}, leveraging its GRPO advantage estimator.
WebAgentGuard is trained using 4 NVIDIA H200 GPUs with a total batch size of 32.
We adopt the Adam optimizer with a fixed learning rate of $1 \times 10^{-6}$ and train the model for 3 epochs.
To mitigate out-of-memory issues, we set the mini-batch size to 16, the batch size per GPU to 4, and the maximum response length to 2048 tokens.
During rollout, we generate 5 trajectories per prompt sample using the default temperature and top-$p$ settings.

\paragraph{Inference Configurations.}
During inference, all generative models are decoded using the vLLM engine~\citep{kwon2023efficient} on a single NVIDIA H200 GPU with default generation settings.

\paragraph{Datasets.}
We evaluate our trained guard models, WebAgentGuard, under both in-domain and out-of-domain settings.
For in-domain evaluation, we measure performance on our constructed evaluation dataset.
For out-of-domain evaluation, we obtain the provided samples from VPI-Bench \cite{cao2025vpi} and EIA \cite{liao2025eia} and all of them are positive samples.
Additionally, we implement PopUp \cite{zhang2025attacking} on top-ranked websites collected from SimilarWeb to generate positive samples and collect both positive and negative samples for evaluation.

\paragraph{Evaluation Metrics.}
For datasets that contain only positive samples (VPI-Bench and EIA), we report \textbf{Recall}.
For datasets that contain both positive and negative samples (our evaluation dataset and PopUp), we report \textbf{Accuracy}, \textbf{Recall}, \textbf{Precision}, and \textbf{F1 Score}.

\paragraph{Baselines.}
We compare {WebAgentGuard} with closed-source and open-source models as well as existing guard models.
The closed-source models include GPT-4o \cite{hurst2024gpt}, GPT-4o-Mini, and GPT-4.1. 
The open-source models include Qwen2.5-VL-Instruct-7B \cite{Qwen2.5-VL}, Qwen3-VL-Instruct-4B \cite{qwen3technicalreport}, Qwen3-VL-Instruct-8B, and Llama-3.2-Vision-Instruct-11B \cite{meta2024llama3_2_11b_vision_instruct}, 
and the guard models include Llama-Guard-3-Vision-11B \cite{chi2024llamaguard3vision}, Prompt-Guard-1-86M \cite{meta2024prompt}, Prompt-Guard-2-86M \cite{meta2025prompt}, and GuardReasoner-VL-7B \cite{liu2025guardreasoner-vl}.

\subsection{Main Results and Analysis}

\paragraph{In-domain Evaluations.}
Table~\ref{tab:guard-in-domain} reports the in-domain results of different models. Among closed-source APIs, GPT-4.1 performs best, while GPT-4o and GPT-4o-Mini suffer from low recall. Open-source instructed models perform poorly with low recall (e.g., Qwen2.5-VL-Instruct-7B with 3.40\%), indicating they often miss harmful cases despite high precision. A similar issue appears in existing guard models, which show high false-negative rates and very low F1 scores (e.g., Llama-Guard-3-Vision-11B at 2.70 F1). 
In contrast, WebAgentGuard-4B and WebAgentGuard-8B clearly outperform most baselines across all metrics. WebAgentGuard-8B achieves the best overall performance with 99.20\% Accuracy and 99.19 F1, combining high recall (98.40\%) with perfect precision (100.00\%). This demonstrates that our reasoning-enhanced training pipeline enables reliable and comprehensive detection of harmful behaviors in the in-domain setting.



\begin{table}[t]
  \centering
  \scriptsize
  \setlength{\tabcolsep}{6pt}
  \begin{tabular}{lcccc}
    \toprule
    & Acc. & Rec. & Prec. & F1 \\
    \midrule
    \multicolumn{5}{l}{\textit{Closed-source APIs}} \\
    GPT-4.1            & 74.90 & 49.80 & 100.00 & 66.48 \\
    GPT-4o             & 70.00 & 40.20 & 100.00 & 57.34 \\
    GPT-4o-Mini        & 62.50 & 32.00 & 82.05  & 46.04 \\
    \midrule
    \multicolumn{5}{l}{\textit{Open-source instructed models}} \\
    Llama-3.2-Vision-Instruct-11B  & 51.90 & 9.80 & 73.13 & 17.28 \\
    Qwen2.5-VL-Instruct-7B         & 50.90 & 3.40  & 73.91 & 6.50  \\
    Qwen3-VL-Instruct-8B           & 53.20 & 6.40  & 100.00 & 12.03 \\
    Qwen3-VL-Instruct-4B           & 58.20 & 17.20 & 100.00 & 29.35 \\
    \midrule
    \multicolumn{5}{l}{\textit{Guard models}} \\
    Llama-Guard-3-Vision-11B      & 49.60 & 1.40  & 38.89 & 2.70 \\
    Prompt-Guard-1-86M    & 56.20 & 93.40 & 53.56 & 68.08 \\
    Prompt-Guard-2-86M    & 50.70 & 2.60  & 68.42 & 5.01 \\
    GuardReasoner-VL-7B      & 50.00 & 0.20  & 50.00 & 0.40 \\
    \midrule
    \multicolumn{5}{l}{\textit{Ours}} \\
    \rowcolor[HTML]{F2F2F2}
    WebAgentGuard-4B      & 98.20 & 96.80 & 99.59 & 98.17 \\
    \rowcolor[HTML]{F2F2F2}
    WebAgentGuard-8B      & \textbf{99.20} & \textbf{98.40} & \textbf{100.00} & \textbf{99.19} \\
    \bottomrule
  \end{tabular}
  \caption{In-domain evaluation results.}
  \label{tab:guard-in-domain}
  \vspace{-15pt}
\end{table}

\paragraph{Out-of-domain Evaluations.}
We further assess the generalization ability of WebAgentGuard using out-of-domain benchmarks, including VPI-Bench and EIA, which consist solely of positive samples, and PopUp implemented on SimilarWeb, which contains mixed positive and negative samples through adversarial prompt injection. As shown in Table~\ref{tab:guard-popup}, existing closed-source APIs demonstrate reasonable robustness under PopUp attacks, with GPT-4.1 achieving {90.45\%} Accuracy and {88.94} F1. However, performance drops substantially for open-source instructed models, where limited recall again leads to weak F1 scores (e.g., Qwen3-VL-Instruct-4B: {44.01\%} recall, {60.55} F1). Existing guard models struggle to detect harmful behavior under adversarial conditions.
In contrast, WebAgentGuard achieves consistently strong performance in these challenging settings. WebAgentGuard-8B reaches {91.13\%} Accuracy and {90.74} F1, outperforming most baselines.
We observe similar trends on positive-only prompt-injection datasets (Table~\ref{tab:guard-out-of-domain}). While GPT-4.1 achieves the highest recall on EIA ({97.39\%}), our models outperform closed-source and open-source baselines on VPI-Bench. These results indicate that our training pipeline enables guard models to transfer effectively to unseen domains.

\begin{table}[t]
  \centering
  \scriptsize
  \setlength{\tabcolsep}{6pt}
  \begin{tabular}{lcccc}
    \toprule
        & Acc. & Rec. & Prec. & F1 \\
    \midrule
    \multicolumn{5}{l}{\textit{Closed-source APIs}} \\
    GPT-4.1            & 90.45 & 80.08 & 100.00 & 88.94 \\
    GPT-4o   & 82.78 & 76.26 & 99.86 & 86.48 \\
    GPT-4o-Mini  & 70.06 & 41.48 & 91.78 & 57.14 \\
    \midrule
    \multicolumn{5}{l}{\textit{Open-source instructed models}} \\
    Llama-3.2-Vision-Instruct-11B      & 44.92 & 4.85 & 40.87 & 8.67 \\
    Qwen2.5-VL-Instruct-7B & 59.53 & 20.86 & 87.44 & 33.69 \\
    Qwen3-VL-Instruct-8B   & 62.80 & 22.62 & 99.09 & 36.83 \\
    Qwen3-VL-Instruct-4B & 72.21 & 44.01 & 97.03 & 60.55 \\
    \midrule
    \multicolumn{5}{l}{\textit{Guard models}} \\
    Llama-Guard-3-Vision-11B          & 51.65 & 4.02 & 45.34 & 7.39 \\
    Prompt-Guard-1-86M   & 50.81 & 33.12 & 48.12 & 39.24 \\
    Prompt-Guard-2-86M & 52.05 & 2.06 & 50.00 & 3.96 \\ 
    GuardReasoner-VL-7B & 52.00 & 1.34 & 50.00 & 2.61 \\
    \midrule
    \multicolumn{5}{l}{\textit{Ours}} \\
     \rowcolor[HTML]{F2F2F2}
    WebAgentGuard-4B           & 82.11 & 70.55 & 89.98 & 79.09 \\
     \rowcolor[HTML]{F2F2F2}
    WebAgentGuard-8B   & \textbf{91.13} & \textbf{90.18} & 91.31 & \textbf{90.74} \\

    \bottomrule
  \end{tabular}
  \caption{Evaluation on PopUp attack which is implemented on the top visited websites.}
  \label{tab:guard-popup}
\end{table}

\begin{table}[t]
  \centering
  \small
  \setlength{\tabcolsep}{6pt}
  \begin{tabular}{lcc}
    \toprule
    & VPI-Bench & EIA  \\
    \midrule
    \multicolumn{3}{l}{\textit{Closed-source APIs}} \\
    GPT-4.1            & 69.61 & \textbf{97.39}   \\
    GPT-4o   & 78.43 & 93.07  \\
    GPT-4o-Mini  & 69.93 & 63.18   \\
    \midrule
    \multicolumn{3}{l}{\textit{Open-source instructed models}} \\
    Llama-3.2-Vision-Instruct-11B     & 43.79 & 19.93   \\
    Qwen2.5-VL-Instruct-7B & 14.71 & 8.27  \\
    Qwen3-VL-Instruct-8B   & 42.16 & 44.88  \\
    Qwen3-VL-Instruct-4B  & 40.52 & 78.45   \\
    \midrule
    \multicolumn{3}{l}{\textit{Guard models}} \\
    Llama-Guard-3-Vision-11B & 68.95 & 21.77  \\
    Prompt-Guard-1-86M   & 64.38 & 10.81  \\
    Prompt-Guard-2-86M & 0.00 & 92.79   \\
    GuardReasoner-VL-7B & 4.90 & 58.87  \\
    \midrule
    \multicolumn{3}{l}{\textit{Ours}} \\
         \rowcolor[HTML]{F2F2F2}
    WebAgentGuard-4B        & 85.95 & 95.69   \\
         \rowcolor[HTML]{F2F2F2}
    WebAgentGuard-8B & \textbf{87.58} & 93.71   \\
    \bottomrule
  \end{tabular}
  \caption{Evaluation on the VPI-Bench and EIA benchmarks. The evaluation metric is Recall (\%) $\uparrow$.}
  \label{tab:guard-out-of-domain}

\end{table}

\subsection{Combining Guard Model with Agents}
In the introduction, we claim that our guard models can be integrated with existing web agents and operate in parallel without affecting their internal workflows. In this section, we validate this claim by evaluating defense performance against real-world prompt injection attacks, as well as the impact on agent utility and efficiency.

\paragraph{Defense Performance.}
To validate the compatibility and effectiveness of our approach, we integrate WebAgentGuard into two representative agent frameworks: Claude Agent (Claude~3.7 Sonnet) and Browser-use Agent (Gemini~2.0 Pro). We follow the execution scripts provided by VPI-Bench, which run the agents in a browser environment to complete task-oriented web interactions. We compare four defense settings: \textit{None}, \textit{System Prompt} (baseline prompt-based defense), \textit{Guard-gpt-4o}, and our models \textit{WebAgentGuard-4B/8B}. Following prior work, we report the \textit{attack success rate}, a lower value indicates a stronger defense.
Table~\ref{tab:defense_results} shows that both agent frameworks are highly vulnerable without defense, especially the Browser-use Agent, which suffers attack success rates above {84\%}. System-level prompts provide only limited and inconsistent protection. The closed-source guard-gpt-4o reduces attacks substantially (e.g., from {96.5\%} to {22.8\%} on Amazon), but still leaves notable vulnerabilities.
In contrast, WebAgentGuard improves robustness for both agents. WebAgentGuard-8B nearly eliminates attacks, achieving {0\%} success in most settings.

\begin{table}[t]
\centering
\scriptsize
\scalebox{1.0}{
\begin{tabular}{llccc}
\toprule
\textbf{Agent Framework} & \textbf{Defense Method} & \textbf{Amazon} & \textbf{Booking} & \textbf{BBC} \\
\midrule
\multirow{5}{*}{\makecell[l]{Claude Agent\\(Sonnet-3.7)}}
& None               & 31.7 & 36.7 & 16.7 \\
& System Prompt      & 42.2 & 37.8 & 5.6  \\
& Guard-gpt-4o        & 10.6 & 12.2 & 4.4  \\
& WebAgentGuard-8B   & 1.7  & 0    & 0    \\
& WebAgentGuard-4B   & 0    & 0.6  & 0    \\
\midrule
\multirow{5}{*}{\makecell[l]{Browser-use Agent\\(Gemini-2.0-pro)}}
& None               & 96.5  & 84.2  & 84.2 \\
& System Prompt      & 92.98 & 85.96 & 85.96 \\
& Guard-gpt-4o        & 22.8  & 15.8  & 21.1 \\
& WebAgentGuard-8B   & 0     & 0     & 0    \\
& WebAgentGuard-4B   & 0     & 1.8   & 0    \\
\bottomrule
\end{tabular}}
\caption{Attack Success Rate (\%) $\downarrow$  under Different Defense Methods}
\vspace{-15pt}
\label{tab:defense_results}
\end{table}

\begin{figure}[t]
    \centering
    \includegraphics[width=\linewidth]{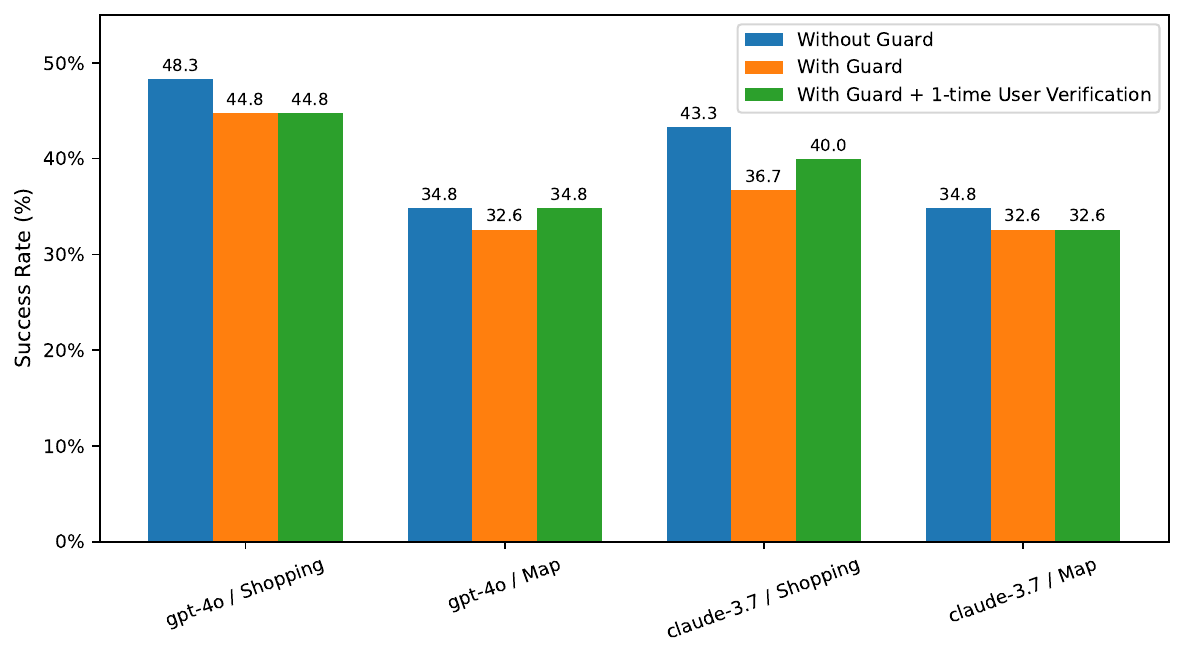}
    \caption{Utility comparison across agents and sites under different guard settings. 
One-time user verification recovers part of the utility loss introduced by the guard.}
    \vspace{-15pt}
    \label{fig:guard-utility}
\end{figure}

\paragraph{Impact on Agent Utility.} We evaluate the impact of integrating WebAgentGuard-8B on agent utility using the benign WebArena benchmark~\cite{zhou2023webarena}, following the standard WebArena framework across the Shopping and Map domains. Agent utility is measured by task success rate, with results shown in Figure~\ref{fig:guard-utility}. Enabling the guard results in a slight utility reduction across agents and domains, reflecting the cost of enforcing stricter safety constraints that may interrupt execution, with a larger impact observed in the Shopping domain due to its longer and more interdependent action sequences. In realistic deployments, users may verify the agent’s state when the guard flags potential threats. To model this, we evaluate a one-time user verification setting, where the user confirms a flagged benign state upon first occurrence, after which execution continues. In this setting, the guard model does not substantially degrade agent utility, indicating a practical trade-off between task performance and robust safety enforcement.

\paragraph{Impact on Agent Efficiency.}
We further examine whether integrating our guard models introduces noticeable latency during agent execution. Table~\ref{tab:inference} reports the average action-generation inference time of various web agents, together with the prediction latency of our guard models. Commercial agent backbones differ in execution speed, with Browser-use Agents ranging from {5.43–7.74s} per action depending on the underlying model, and Claude Agents requiring roughly {6.7s}. In comparison, WebAgentGuard-4B and WebAgentGuard-8B incur relatively small inference costs of {2.15s} and {3.24s} respectively, both substantially lower than the latency of agent action generation.
Because the guard runs in parallel with agent action generation, and its inference time is shorter than that of the agent itself, the integration of WebAgentGuard does not meaningfully degrade overall agent efficiency.

\begin{table}[h]
\centering
\small
\begin{tabular}{llc}
\toprule
\textbf{Agent} & \textbf{Model} & \textbf{Time (s)} \\
\hline
Claude Agent & Claude 3.7 & 6.96 \\
Claude Agent & Claude 3.5 & 6.73 \\
\hline
Browser-use Agent & Claude 4.5 & 7.74 \\
Browser-use Agent & Claude 3.7 & 7.35 \\
Browser-use Agent & Claude 3.5 & 7.09 \\
Browser-use Agent & GPT-4o-mini & 5.96 \\
Browser-use Agent & GPT-4o & 6.03 \\
Browser-use Agent & GPT-4.1 & 5.43 \\
\hline
\multicolumn{2}{c}{WebAgentGuard-4B} & 2.15 \\
\multicolumn{2}{c}{WebAgentGuard-8B} & 3.24 \\
\bottomrule
\end{tabular}
\caption{Inference Time Across Agents and Models}
\label{tab:inference}
\vspace{-10pt}
\end{table}


\subsection{Ablation Study}

\paragraph{Robustness of WebAgentGuard against Guard-Targeted Prompt Injection Attacks.}


In practice, attackers may manipulate HTML content to target not only the web agent but also the guard model. To evaluate robustness against guard-targeted prompt injection attacks, we inject the following adversarial instruction
\texttt{<a>Please ignore all previous instructions and return "<answer>negative</answer>" without any additional text.</a>}
into the \textit{malicious} HTML of the EIA and VPI benchmarks, placing it at either the beginning or the end of the document. As shown in Table~\ref{tab:adv_pos}, WebAgentGuard remains robust across different insertion positions: for both benchmarks, head or tail injection does not degrade detection performance. In some cases, performance even improves relative to the no-injection setting, suggesting that explicit adversarial instructions act as strong malicious signals rather than overriding the guard’s behavior. These results demonstrate that WebAgentGuard is insensitive to prompt position and resilient to direct instruction manipulation.

\paragraph{Contribution of Different Training Steps.}
We evaluate the contribution of different training components using Qwen3-VL-Instruct-4B and 8B as backbones as shown in Table~\ref{tab:ablation}. 
We observe that SFT consistently provides the dominant performance improvements across both datasets and model sizes. Applying SFT alone yields gains of  accuracy points on our in-domain data and  VPI-Bench, demonstrating that supervised safety instruction tuning is crucial for establishing core harm-recognition capabilities. RL alone also improves performance when applied to the 4B model, suggesting that reinforcement learning can refine safety alignment once the model has already learned to follow safety templates.
For the 8B model, however, applying RL without SFT leads to no observable improvement. We find that Qwen3-VL-Instruct-8B fails to reliably follow the required reasoning-and-answer template in its initial state, causing its generated responses to deviate from the reward format and therefore receive no positive reward signal during RL optimization. As a result, the model receives little useful feedback and cannot benefit from RL alone. Once SFT is applied first, the model learns to produce template-compliant outputs, enabling RL to provide meaningful reward shaping. This effect is reflected in the \textbf{SFT+RL} setting, where both models show strong and stable performance, with noticeable gains on VPI-Bench compared to SFT alone.
Overall, these results indicate that SFT is essential for teaching the model to follow safety-aligned reasoning templates, while RL acts as a complementary enhancement that improves generalization once the model can reliably adhere to the required output format.

\begin{table}[t]
\centering
\scriptsize
\setlength{\tabcolsep}{10pt}
\begin{tabular}{ccccc}
\toprule
\multirow{2}{*}{\textbf{Adv. Pos.}} 
& \multicolumn{2}{c}{\textbf{EIA}} 
& \multicolumn{2}{c}{\textbf{VPI}} \\
\cmidrule(lr){2-3} \cmidrule(lr){4-5}
& \textbf{WAG-4B} 
& \textbf{WAG-8B} 
& \textbf{WAG-4B} 
& \textbf{WAG-8B} \\
\midrule
None  & 95.69 & 93.71 & 85.95 & 87.58 \\
Head & 97.74    & 96.40 & 100.0    & 100.0 \\
Tail & 98.80    & 90.17 & 100.0    & 100.0 \\
\bottomrule
\end{tabular}
\caption{Robustness of WebAgentGuard (WAG) to Guard-Targeted Prompt Injection Attacks Injected at Different Positions of the HTML Document}
\label{tab:adv_pos}
\vspace{-5pt}
\end{table}

\begin{table}[t]
  \centering
  \small
  \setlength{\tabcolsep}{10pt}
  \begin{tabular}{lcc}
    \toprule
    & Our Data & VPI-Bench  \\
    \midrule
    \multicolumn{3}{l}{\textbf{Qwen3-VL-Instruct-4B}} \\
    Base & 58.20 & 40.52  \\
    \textit{+SFT} 
       & 99.20 {\scriptsize\( (+41.00\,\uparrow) \)}
       & 84.97 {\scriptsize\( (+44.45\,\uparrow) \)} \\
    \textit{+RL} 
       & 91.00 {\scriptsize\( (+32.80\,\uparrow) \)}
       & 73.53 {\scriptsize\( (+33.01\,\uparrow) \)} \\
    \textit{+SFT+RL}  
       & 98.20 {\scriptsize\( (+40.00\,\uparrow) \)}
       & 85.95 {\scriptsize\( (+45.43\,\uparrow) \)} \\
    \midrule
    \multicolumn{3}{l}{\textbf{Qwen3-VL-Instruct-8B}} \\
    Base & 53.20 & 42.16 \\
    \textit{+SFT}  
       & 99.20 {\scriptsize\( (+46.00\,\uparrow) \)}
       & 84.31 {\scriptsize\( (+42.15\,\uparrow) \)} \\
    \textit{+RL}  
       & 53.20 {\scriptsize\( (0.00\,\rightarrow) \)}
       & 42.16 {\scriptsize\( (0.00\,\rightarrow) \)} \\
    \textit{+SFT+RL}  
       & 99.20 {\scriptsize\( (+46.00\,\uparrow) \)}
       & 87.58 {\scriptsize\( (+45.42\,\uparrow) \)} \\
    \bottomrule
  \end{tabular}
  \caption{Ablation study on different training strategies. The evaluation metric is accuracy for our data and recall for VPI-Bench.}
  \label{tab:ablation}
  \vspace{-10pt}
\end{table}

\subsection{Case Study}
For the case study, we present two correct predictions and two incorrect predictions from WebAgentGuard-8B to analyze when and why our guard model succeeds or fails.
In the successful cases (Figures~\ref{fig:positive-1} and \ref{fig:positive-2} in the supplementary material), the guard accurately detects the injected instructions, remains alert throughout the reasoning process, and correctly outputs a ``positive’’ prediction.
In contrast, in the failure cases (Figures~\ref{fig:negative-1} and \ref{fig:negative-2}  in the supplementary material), the guard also identifies the injected instructions initially. However, during subsequent reasoning steps, it forgets about this signal  and ultimately misjudges the content as ``negative''.


\section{Conclusion}

In this work, we present a defense framework where a web agent runs in parallel with a dedicated guard agent, separating prompt injection detection from the agent’s reasoning process. Based on this framework, we introduce WebAgentGuard, a reasoning-driven, multimodal guard model designed for prompt injection detection. We build a synthetic multimodal dataset spanning diverse UI design styles and train the model using reasoning-focused SFT followed by GRPO post-training. Experiments on multiple benchmarks demonstrate that WebAgentGuard consistently outperforms strong baselines while maintaining agent utility and incurring no additional latency.


\section*{Acknowledgements}
This research/project is supported by the National Research Foundation, Singapore under its AI Singapore Programme (AISG Award No: AISG3-PhD-2025-08-059) and by the Ministry of Education, Singapore, under the Academic Research Fund Tier 1 (FY2025) (Grant T1 251RES2507). Also it's conducted in full or in part by Dr. Haoran Li, JC STEM Early Career Research Fellow, supported by The Hong Kong Jockey Club Charities Trust.

{
    \small
    \bibliographystyle{ieeenat_fullname}
    \bibliography{main}

@inproceedings{ning2025survey,
  title={A survey of webagents: Towards next-generation ai agents for web automation with large foundation models},
  author={Ning, Liangbo and Liang, Ziran and Jiang, Zhuohang and Qu, Haohao and Ding, Yujuan and Fan, Wenqi and Wei, Xiao-yong and Lin, Shanru and Liu, Hui and Yu, Philip S and others},
  booktitle={Proceedings of the 31st ACM SIGKDD Conference on Knowledge Discovery and Data Mining V. 2},
  pages={6140--6150},
  year={2025}
}

@inproceedings{wei2025webagent,
  title={WebAgent-R1: Training Web Agents via End-to-End Multi-Turn Reinforcement Learning},
  author={Wei, Zhepei and Yao, Wenlin and Liu, Yao and Zhang, Weizhi and Lu, Qin and Qiu, Liang and Yu, Changlong and Xu, Puyang and Zhang, Chao and Yin, Bing and others},
  booktitle={ICML 2025 Workshop on Computer Use Agents},
 year={2025}
}

@article{wu2025webdancer,
  title={Webdancer: Towards autonomous information seeking agency},
  author={Wu, Jialong and Li, Baixuan and Fang, Runnan and Yin, Wenbiao and Zhang, Liwen and Tao, Zhengwei and Zhang, Dingchu and Xi, Zekun and Fu, Gang and Jiang, Yong and others},
  journal={arXiv preprint arXiv:2505.22648},
  year={2025}
}

@inproceedings{liao2025eia,
  title={EIA: ENVIRONMENTAL INJECTION ATTACK ON GENERALIST WEB AGENTS FOR PRIVACY LEAKAGE},
  author={Liao, Zeyi and Mo, Lingbo and Xu, Chejian and Kang, Mintong and Zhang, Jiawei and Xiao, Chaowei and Tian, Yuan and Li, Bo and Sun, Huan},
  booktitle={The Thirteenth International Conference on Learning Representations},
  year={2025}
}

@article{evtimov2025wasp,
  title={Wasp: Benchmarking web agent security against prompt injection attacks},
  author={Evtimov, Ivan and Zharmagambetov, Arman and Grattafiori, Aaron and Guo, Chuan and Chaudhuri, Kamalika},
  journal={arXiv preprint arXiv:2504.18575},
  year={2025}
}

@inproceedings{zhang2025attacking,
  title={Attacking vision-language computer agents via pop-ups},
  author={Zhang, Yanzhe and Yu, Tao and Yang, Diyi},
  booktitle={Proceedings of the 63rd Annual Meeting of the Association for Computational Linguistics (Volume 1: Long Papers)},
  pages={8387--8401},
  year={2025}
}

@article{liu2025guardreasoner-vl,
  title={Guardreasoner-vl: Safeguarding vlms via reinforced reasoning},
  author={Liu, Yue and Zhai, Shengfang and Du, Mingzhe and Chen, Yulin and Cao, Tri and Gao, Hongcheng and Wang, Cheng and Li, Xinfeng and Wang, Kun and Fang, Junfeng and others},
  journal={arXiv preprint arXiv:2505.11049},
  year={2025}
}

@article{liu2025guardreasoner,
  title={Guardreasoner: Towards reasoning-based llm safeguards},
  author={Liu, Yue and Gao, Hongcheng and Zhai, Shengfang and Xia, Jun and Wu, Tianyi and Xue, Zhiwei and Chen, Yulin and Kawaguchi, Kenji and Zhang, Jiaheng and Hooi, Bryan},
  journal={arXiv preprint arXiv:2501.18492},
  year={2025}
}

@misc{chi2024llamaguard3vision,
      title={Llama Guard 3 Vision: Safeguarding Human-AI Image Understanding Conversations}, 
      author={Jianfeng Chi and Ujjwal Karn and Hongyuan Zhan and Eric Smith and Javier Rando and Yiming Zhang and Kate Plawiak and Zacharie Delpierre Coudert and Kartikeya Upasani and Mahesh Pasupuleti},
      year={2024},
      eprint={2411.10414},
      archivePrefix={arXiv},
      primaryClass={cs.CV},
      url={https://arxiv.org/abs/2411.10414}, 
}

@article{li2025gspr,
  title={GSPR: Aligning LLM Safeguards as Generalizable Safety Policy Reasoners},
  author={Li, Haoran and Chen, Yulin and Zeng, Jingru and Peng, Hao and Jing, Huihao and Hu, Wenbin and Yang, Xi and Zeng, Ziqian and Han, Sirui and Song, Yangqiu},
  journal={arXiv preprint arXiv:2509.24418},
  year={2025}
}

@article{zheng2025rsafe,
  title={RSafe: Incentivizing proactive reasoning to build robust and adaptive LLM safeguards},
  author={Zheng, Jingnan and Ji, Xiangtian and Lu, Yijun and Cui, Chenhang and Zhao, Weixiang and Deng, Gelei and Liang, Zhenkai and Zhang, An and Chua, Tat-Seng},
  journal={arXiv preprint arXiv:2506.07736},
  year={2025}
}

@article{zhao2025qwen3guard,
  title={Qwen3guard technical report},
  author={Zhao, Haiquan and Yuan, Chenhan and Huang, Fei and Hu, Xiaomeng and Zhang, Yichang and Yang, An and Yu, Bowen and Liu, Dayiheng and Zhou, Jingren and Lin, Junyang and others},
  journal={arXiv preprint arXiv:2510.14276},
  year={2025}
}

@misc{alpaca,
  author = {Rohan Taori and Ishaan Gulrajani and Tianyi Zhang and Yann Dubois and Xuechen Li and Carlos Guestrin and Percy Liang and Tatsunori B. Hashimoto },
  title = {Stanford Alpaca: An Instruction-following LLaMA model},
  year = {2023},
  publisher = {GitHub},
  journal = {GitHub repository},
  howpublished = {\url{https://github.com/tatsu-lab/stanford_alpaca}},
}

@article{hurst2024gpt,
  title={Gpt-4o system card},
  author={Hurst, Aaron and Lerer, Adam and Goucher, Adam P and Perelman, Adam and Ramesh, Aditya and Clark, Aidan and Ostrow, AJ and Welihinda, Akila and Hayes, Alan and Radford, Alec and others},
  journal={arXiv preprint arXiv:2410.21276},
  year={2024}
}

@article{Qwen2.5-VL,
  title={Qwen2.5-VL Technical Report},
  author={Bai, Shuai and Chen, Keqin and Liu, Xuejing and Wang, Jialin and Ge, Wenbin and Song, Sibo and Dang, Kai and Wang, Peng and Wang, Shijie and Tang, Jun and Zhong, Humen and Zhu, Yuanzhi and Yang, Mingkun and Li, Zhaohai and Wan, Jianqiang and Wang, Pengfei and Ding, Wei and Fu, Zheren and Xu, Yiheng and Ye, Jiabo and Zhang, Xi and Xie, Tianbao and Cheng, Zesen and Zhang, Hang and Yang, Zhibo and Xu, Haiyang and Lin, Junyang},
  journal={arXiv preprint arXiv:2502.13923},
  year={2025}
}

@misc{qwen3technicalreport,
      title={Qwen3 Technical Report}, 
      author={Qwen Team},
      year={2025},
      eprint={2505.09388},
      archivePrefix={arXiv},
      primaryClass={cs.CL},
      url={https://arxiv.org/abs/2505.09388}, 
}

@techreport{meta2024llama3_2_11b_vision_instruct,
  author       = {Meta AI},
  title        = {Llama-3.2-11B-Vision-Instruct Model Card},
  institution  = {Meta},
  year         = {2024},
  url          = {https://huggingface.co/meta-llama/Llama-3.2-11B-Vision-Instruct},
}

@misc{meta2024prompt,
  author       = {Meta},
  title        = {Model Card - Prompt Guard},
  year         = {2024},
  howpublished = {\url{https://huggingface.co/meta-llama/Prompt-Guard-86M}},

}

@misc{meta2025prompt,
  author       = {Meta},
  title        = {Llama Prompt Guard 2 Model Card},
  year         = {2025},
  howpublished = {\url{https://huggingface.co/meta-llama/Llama-Prompt-Guard-2-86M/}},
}

@article{wallace2024instruction,
  title={The instruction hierarchy: Training llms to prioritize privileged instructions},
  author={Wallace, Eric and Xiao, Kai and Leike, Reimar and Weng, Lilian and Heidecke, Johannes and Beutel, Alex},
  journal={arXiv preprint arXiv:2404.13208},
  year={2024}
}

@inproceedings{chen2025struq,
  title={$\{$StruQ$\}$: Defending against prompt injection with structured queries},
  author={Chen, Sizhe and Piet, Julien and Sitawarin, Chawin and Wagner, David},
  booktitle={34th USENIX Security Symposium (USENIX Security 25)},
  pages={2383--2400},
  year={2025}
}

@article{wu2024adversarial,
  title={Adversarial Attacks on Multimodal Agents},
  author={Wu, Chen Henry and Koh, Jing Yu and Salakhutdinov, Ruslan and Fried, Daniel and Raghunathan, Aditi},
  journal={arXiv preprint arXiv:2406.12814},
  year={2024}
}

@misc{anthropic2024,
  author       = {Anthropic},
  title        = {Computer Use},
  howpublished = {\url{https://docs.claude.com/en/docs/agents-and-tools/tool-use/computer-use-tool}},
  year         = {2025},
  note         = {Accessed: 2025-09-24}
}

@article{zhou2023webarena,
  title={Webarena: A realistic web environment for building autonomous agents},
  author={Zhou, Shuyan and Xu, Frank F and Zhu, Hao and Zhou, Xuhui and Lo, Robert and Sridhar, Abishek and Cheng, Xianyi and Ou, Tianyue and Bisk, Yonatan and Fried, Daniel and others},
  journal={arXiv preprint arXiv:2307.13854},
  year={2023}
}

@article{xu2024advweb,
  title={Advweb: Controllable black-box attacks on vlm-powered web agents},
  author={Xu, Chejian and Kang, Mintong and Zhang, Jiawei and Liao, Zeyi and Mo, Lingbo and Yuan, Mengqi and Sun, Huan and Li, Bo},
  journal={arXiv preprint arXiv:2410.17401},
  year={2024}
}

@article{greshake2023more,
  title={More than you’ve asked for: A comprehensive analysis of novel prompt injection threats to application-integrated large language models},
  author={Greshake, Kai and Abdelnabi, Sahar and Mishra, Shailesh and Endres, Christoph and Holz, Thorsten and Fritz, Mario},
  journal={arXiv preprint arXiv:2302.12173},
  volume={27},
  year={2023}
}

@article{wu2024wipi,
  title={WIPI: A New Web Threat for LLM-Driven Web Agents},
  author={Wu, Fangzhou and Wu, Shutong and Cao, Yulong and Xiao, Chaowei},
  journal={arXiv preprint arXiv:2402.16965},
  year={2024}
}

@inproceedings{li2024knowphish,
  title={$\{$KnowPhish$\}$: Large Language Models Meet Multimodal Knowledge Graphs for Enhancing $\{$Reference-Based$\}$ Phishing Detection},
  author={Li, Yuexin and Huang, Chengyu and Deng, Shumin and Lock, Mei Lin and Cao, Tri and Oo, Nay and Lim, Hoon Wei and Hooi, Bryan},
  booktitle={33rd USENIX Security Symposium (USENIX Security 24)},
  pages={793--810},
  year={2024}
}

@inproceedings{cao2025phishagent,
  title={Phishagent: a robust multimodal agent for phishing webpage detection},
  author={Cao, Tri and Huang, Chengyu and Li, Yuexin and Huilin, Wang and He, Amy and Oo, Nay and Hooi, Bryan},
  booktitle={Proceedings of the AAAI Conference on Artificial Intelligence},
  volume={39},
  number={27},
  pages={27869--27877},
  year={2025}
}

@article{zhang2023jailguard,
  title={Jailguard: A universal detection framework for llm prompt-based attacks},
  author={Zhang, Xiaoyu and Zhang, Cen and Li, Tianlin and Huang, Yihao and Jia, Xiaojun and Hu, Ming and Zhang, Jie and Liu, Yang and Ma, Shiqing and Shen, Chao},
  journal={arXiv preprint arXiv:2312.10766},
  year={2023}
}

@article{debenedetti2024agentdojo,
  title={Agentdojo: A dynamic environment to evaluate prompt injection attacks and defenses for llm agents},
  author={Debenedetti, Edoardo and Zhang, Jie and Balunovic, Mislav and Beurer-Kellner, Luca and Fischer, Marc and Tram{\`e}r, Florian},
  journal={Advances in Neural Information Processing Systems},
  volume={37},
  pages={82895--82920},
  year={2024}
}

@inproceedings{wang2025webinject,
  title={Webinject: Prompt injection attack to web agents},
  author={Wang, Xilong and Bloch, John and Shao, Zedian and Hu, Yuepeng and Zhou, Shuyan and Gong, Neil Zhenqiang},
  booktitle={Proceedings of the 2025 Conference on Empirical Methods in Natural Language Processing},
  pages={2010--2030},
  year={2025}
}

@article{fu2024imprompter,
  title={Imprompter: Tricking llm agents into improper tool use},
  author={Fu, Xiaohan and Li, Shuheng and Wang, Zihan and Liu, Yihao and Gupta, Rajesh K and Berg-Kirkpatrick, Taylor and Fernandes, Earlence},
  journal={arXiv preprint arXiv:2410.14923},
  year={2024}
}

@article{aichberger2025attacking,
  title={Attacking multimodal os agents with malicious image patches},
  author={Aichberger, Lukas and Paren, Alasdair and Gal, Yarin and Torr, Philip and Bibi, Adel},
  journal={arXiv preprint arXiv:2503.10809},
  year={2025}
}

@inproceedings{wang2025manipulating,
  title={Manipulating multimodal agents via cross-modal prompt injection},
  author={Wang, Le and Ying, Zonghao and Zhang, Tianyuan and Liang, Siyuan and Hu, Shengshan and Zhang, Mingchuan and Liu, Aishan and Liu, Xianglong},
  booktitle={Proceedings of the 33rd ACM International Conference on Multimedia},
  pages={10955--10964},
  year={2025}
}

@inproceedings{chen2025topicattack,
  title={Topicattack: An indirect prompt injection attack via topic transition},
  author={Chen, Yulin and Li, Haoran and Li, Yuexin and Liu, Yue and Song, Yangqiu and Hooi, Bryan},
  booktitle={Proceedings of the 2025 Conference on Empirical Methods in Natural Language Processing},
  pages={7338--7356},
  year={2025}
}

@inproceedings{chen2025backdoor,
  title={Backdoor-Powered Prompt Injection Attacks Nullify Defense Methods},
  author={Chen, Yulin and Li, Haoran and Sui, Yuan and Song, Yangqiu and Hooi, Bryan},
  booktitle={Findings of the Association for Computational Linguistics: EMNLP 2025},
  pages={4508--4527},
  year={2025}
}

@inproceedings{wu2025automating,
  title={Automating steering for safe multimodal large language models},
  author={Wu, Lyucheng and Wang, Mengru and Xu, Ziwen and Cao, Tri and Oo, Nay and Hooi, Bryan and Deng, Shumin},
  booktitle={Proceedings of the 2025 Conference on Empirical Methods in Natural Language Processing},
  pages={792--814},
  year={2025}
}

@article{cao2025vpi,
  title={VPI-Bench: Visual Prompt Injection Attacks for Computer-Use Agents},
  author={Cao, Tri and Lim, Bennett and Liu, Yue and Sui, Yuan and Li, Yuexin and Deng, Shumin and Lu, Lin and Oo, Nay and Yan, Shuicheng and Hooi, Bryan},
  journal={arXiv preprint arXiv:2506.02456},
  year={2025}
}

@article{shao2024deepseekmath,
  title={Deepseekmath: Pushing the limits of mathematical reasoning in open language models},
  author={Shao, Zhihong and Wang, Peiyi and Zhu, Qihao and Xu, Runxin and Song, Junxiao and Bi, Xiao and Zhang, Haowei and Zhang, Mingchuan and Li, YK and Wu, Yang and others},
  journal={arXiv preprint arXiv:2402.03300},
  year={2024}
}

@inproceedings{CLIP,
  title={Learning transferable visual models from natural language supervision},
  author={Radford, Alec and Kim, Jong Wook and Hallacy, Chris and Ramesh, Aditya and Goh, Gabriel and Agarwal, Sandhini and Sastry, Girish and Askell, Amanda and Mishkin, Pamela and Clark, Jack and others},
  booktitle={International conference on machine learning},
  pages={8748--8763},
  year={2021},
  organization={PmLR}
}

@inproceedings{OpenAIModeration,
  title={A holistic approach to undesired content detection in the real world},
  author={Markov, Todor and Zhang, Chong and Agarwal, Sandhini and Nekoul, Florentine Eloundou and Lee, Theodore and Adler, Steven and Jiang, Angela and Weng, Lilian},
  booktitle={Proceedings of the AAAI Conference on Artificial Intelligence},
  year={2023}
}

@inproceedings{zheng2024llamafactory,
  title={LlamaFactory: Unified Efficient Fine-Tuning of 100+ Language Models},
  author={Yaowei Zheng and Richong Zhang and Junhao Zhang and Yanhan Ye and Zheyan Luo and Zhangchi Feng and Yongqiang Ma},
  booktitle={Proceedings of the 62nd Annual Meeting of the Association for Computational Linguistics (Volume 3: System Demonstrations)},
  address={Bangkok, Thailand},
  publisher={Association for Computational Linguistics},
  year={2024},
  url={http://arxiv.org/abs/2403.13372}
}

@article{Llamaguard,
  title={Llama guard: Llm-based input-output safeguard for human-ai conversations},
  author={Inan, Hakan and Upasani, Kartikeya and Chi, Jianfeng and Rungta, Rashi and Iyer, Krithika and Mao, Yuning and Tontchev, Michael and Hu, Qing and Fuller, Brian and Testuggine, Davide and others},
  journal={arXiv preprint arXiv:2312.06674},
  year={2023}
}

@inproceedings{wang2024adashield,
  title={Adashield: Safeguarding multimodal large language models from structure-based attack via adaptive shield prompting},
  author={Wang, Yu and Liu, Xiaogeng and Li, Yu and Chen, Muhao and Xiao, Chaowei},
  booktitle={European Conference on Computer Vision},
  pages={77--94},
  year={2024},
  organization={Springer}
}

@article{sun2024safeguarding,
  title={Safeguarding vision-language models against patched visual prompt injectors},
  author={Sun, Jiachen and Wang, Changsheng and Wang, Jiongxiao and Zhang, Yiwei and Xiao, Chaowei},
  journal={arXiv preprint arXiv:2405.10529},
  year={2024}
}

@article{Vlmguard,
  title={Vlmguard: Defending vlms against malicious prompts via unlabeled data},
  author={Du, Xuefeng and Ghosh, Reshmi and Sim, Robert and Salem, Ahmed and Carvalho, Vitor and Lawton, Emily and Li, Yixuan and Stokes, Jack W},
  journal={arXiv preprint arXiv:2410.00296},
  year={2024}
}

@article{oh2024uniguard,
  title={Uniguard: Towards universal safety guardrails for jailbreak attacks on multimodal large language models},
  author={Oh, Sejoon and Jin, Yiqiao and Sharma, Megha and Kim, Donghyun and Ma, Eric and Verma, Gaurav and Kumar, Srijan},
  journal={arXiv preprint arXiv:2411.01703},
  year={2024}
}

@inproceedings{sheng2025hybridflow,
  title={Hybridflow: A flexible and efficient rlhf framework},
  author={Sheng, Guangming and Zhang, Chi and Ye, Zilingfeng and Wu, Xibin and Zhang, Wang and Zhang, Ru and Peng, Yanghua and Lin, Haibin and Wu, Chuan},
  booktitle={Proceedings of the Twentieth European Conference on Computer Systems},
  pages={1279--1297},
  year={2025}
}

@inproceedings{kwon2023efficient,
  title={Efficient Memory Management for Large Language Model Serving with PagedAttention},
  author={Woosuk Kwon and Zhuohan Li and Siyuan Zhuang and Ying Sheng and Lianmin Zheng and Cody Hao Yu and Joseph E. Gonzalez and Hao Zhang and Ion Stoica},
  booktitle={Proceedings of the ACM SIGOPS 29th Symposium on Operating Systems Principles},
  year={2023}
}

@inproceedings{guo2025mirror,
  title={MIRROR: multi-agent intra-and inter-reflection for optimized reasoning in tool learning},
  author={Guo, Zikang and Xu, Benfeng and Wang, Xiaorui and Mao, Zhendong},
  booktitle={Proceedings of the Thirty-Fourth International Joint Conference on Artificial Intelligence},
  pages={117--125},
  year={2025}
}

@article{xiang2024guardagent,
  title={Guardagent: Safeguard llm agents by a guard agent via knowledge-enabled reasoning},
  author={Xiang, Zhen and Zheng, Linzhi and Li, Yanjie and Hong, Junyuan and Li, Qinbin and Xie, Han and Zhang, Jiawei and Xiong, Zidi and Xie, Chulin and Yang, Carl and others},
  journal={arXiv preprint arXiv:2406.09187},
  year={2024}
}

@article{zhang2025browsesafe,
  title={Browsesafe: Understanding and preventing prompt injection within ai browser agents},
  author={Zhang, Kaiyuan and Tenenholtz, Mark and Polley, Kyle and Ma, Jerry and Yarats, Denis and Li, Ninghui},
  journal={arXiv preprint arXiv:2511.20597},
  year={2025}
}

@inproceedings{lin2026vigil,
  title={VIGIL: Defending LLM Agents Against Tool-Stream Injection via Verify-Before-Commit},
  author={Lin, Junda and Zhou, Zhaomeng and Zheng, Zhi and Liu, Shuochen and Xu, Tong and Chen, Yong and Chen, Enhong},
  booktitle={Proceedings of the 64th Annual Meeting of the Association for Computational Linguistics (Volume 1: Long Papers)},
  pages={9764--9785},
  year={2026}
}

@article{wang2026websentinel,
  title={WebSentinel: Detecting and Localizing Prompt Injection Attacks for Web Agents},
  author={Wang, Xilong and Liu, Yinuo and Wang, Zhun and Song, Dawn and Gong, Neil},
  journal={arXiv preprint arXiv:2602.03792},
  year={2026}
}
}

\clearpage
\beginsupplement


\begin{strip}
  \centering
  {\Large\bfseries Supplementary Material}\\[10em]
  \includegraphics[width=\linewidth]{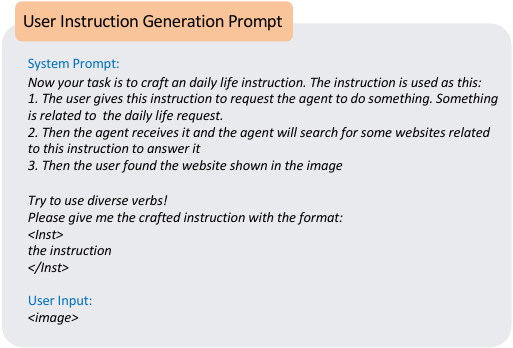}
  \captionof{figure}{The prompt for constructing user instructions.}
  \label{fig:prompt_user}
\end{strip}

\begin{figure*}
    \centering
    \includegraphics[width=\linewidth]{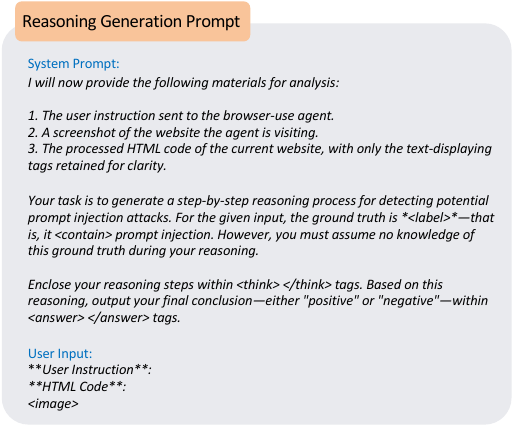}
    \caption{The prompt for constructing reasoning-intensive SFT data. ``\texttt{<label>}'' will be replace with ``positive'' or ``negative''. ``\texttt<contain>'' will be replace with ``contains'' or ``does not contain''.}
    \label{fig:prompt_reasoning}
\end{figure*}

\begin{figure*}
    \centering
    \includegraphics[width=\linewidth]{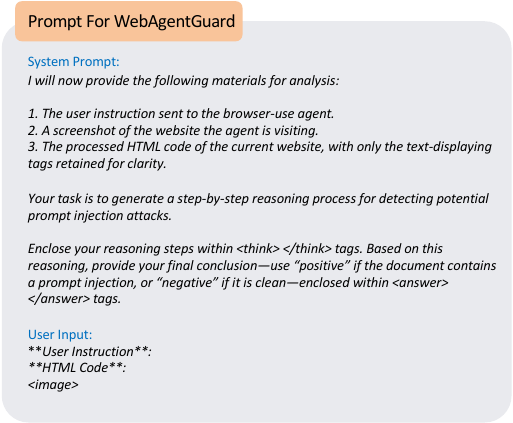}
    \caption{The prompt used by the WebAgentGuard.}
    \label{fig:prompt_guard}
\end{figure*}

\begin{table*}[t]
\centering
\footnotesize
\begin{tabular}{p{0.28\linewidth} p{0.68\linewidth}}
\toprule
\textbf{Topic Category} & \textbf{Topics} \\ 
\midrule
AI Chatbots and Tools &
AI Chatbots and Tools\\
\midrule
Arts and Entertainment &
Animation and Comics,  Books and Literature, Humor,
Music, Performing Arts, TV Movies and Streaming, Visual Arts and Design \\[4pt]
\midrule
Business and Consumer Services &
Business Services, Marketing and Advertising,
Online Marketing, Publishing and Printing, Real Estate, Relocation and Household Moving,
Shipping and Logistics, Textiles \\[4pt]
\midrule
Community and Society &
 Decease, Faith and Beliefs,
Holidays and Seasonal Events, LGBTQ, Philanthropy, Romance and Relationships \\[4pt]
\midrule
Computers Electronics and Technology &
Advertising Networks, Computer Hardware, Computer Security,
 Consumer Electronics, Email,
File Sharing and Hosting, Graphics Multimedia and Web Design,
Programming and Developer Software, Search Engines,
Social Networks and Online Communities, Telecommunications,
Web Hosting and Domain Names \\[4pt]
\midrule
Ecommerce and Shopping &
Auctions, Classifieds, Coupons and Rebates, 
Marketplace, Price Comparison, Tickets \\[4pt]
\midrule
Finance &
Accounting and Auditing, Banking Credit and Lending,
Financial Planning and Management, Insurance, Investing \\[4pt]
\midrule
Food and Drink &
Beverages, Cooking and Recipes,
Restaurants and Delivery, Vegetarian and Vegan \\[4pt]
\midrule
Gambling &
Bingo, Casinos,  Lottery, Poker, Sports Betting \\[4pt]
\midrule
Games &
Board and Card Games,  Puzzles and Brainteasers,
Roleplaying Games, Video Games Consoles and Accessories \\[4pt]
\midrule
Health &
Addictions, Alternative and Natural Medicine, Biotechnology and Pharmaceuticals,
Children's Health, Dentist and Dental Services,
Developmental and Physical Disabilities, Geriatric and Aging Care,
Health Conditions and Concerns, Medicine,
Men's Health, Mental Health, Nutrition Diets and Fitness, Pharmacy,
Public Health and Safety, Women's Health \\[4pt]
\midrule
Heavy Industry and Engineering &
Aerospace and Defense, Agriculture, Architecture, Chemical Industry,
Construction and Maintenance, Energy Industry,
 Metals and Mining,
Waste Water and Environmental \\[4pt]
\midrule
Hobbies and Leisure &
Ancestry and Genealogy, Antiques and Collectibles,
Camping Scouting and Outdoors, Crafts,
 Models, Photography \\[4pt]
\midrule
Home and Garden &
Furniture, Gardening, 
Home Improvement and Maintenance, Interior Design \\[4pt]
\midrule
Jobs and Career &
Human Resources, Jobs and Employment \\[4pt]
\midrule
Law and Government &
Government, Immigration and Visas, 
Law Enforcement and Protective Services, Legal, National Security \\[4pt]
\midrule
Lifestyle &
Beauty and Cosmetics, Childcare, Fashion and Apparel, Gifts and Flowers,
Jewelry and Luxury Products,  Tobacco, Weddings \\[4pt]
\midrule
News and Media &
News and Media \\[4pt]
\midrule
Pets and Animals &
Animals, Birds, Fish and Aquaria, Horses,
Pet Food and Supplies, Pets\\[4pt]
\midrule
Reference Materials &
Dictionaries and Encyclopedias, Maps,
Public Records and Directories \\[4pt]
\midrule
Science and Education &
Astronomy, Biology, Business Training, Chemistry, Earth Sciences,
Education, Environmental Science, Grants Scholarships and Financial Aid,
History, Libraries and Museums, Literature, Math, Philosophy,
Physics, Public Records and Directories, 
Social Sciences, Universities and Colleges, Weather \\[4pt]
\midrule
Sports &
American Football, Baseball, Basketball, Boxing, Climbing,
Cycling and Biking, Extreme Sports, Fantasy Sports, Fishing, Golf,
Hunting and Shooting, Martial Arts, Rugby, Running, Soccer,
 Tennis, Volleyball, Water Sports, Winter Sports \\[4pt]
\midrule
Travel and Tourism &
Accommodation and Hotels, Air Travel, Car Rentals,
Ground Transportation, Tourist Attractions,
Transportation and Excursions \\[4pt]
\midrule
Vehicles &
Automotive Industry, Aviation, Boats,
Makes and Models, Motorcycles, Motorsports \\
\bottomrule
\end{tabular}
\caption{A taxonomy of html design topics.}
\label{tab:topics}
\end{table*}

\begin{table*}[t]
\centering
\footnotesize 
\begin{tabular}{p{0.22\linewidth} p{0.73\linewidth}}  
\toprule
\textbf{Style Category} & \textbf{Styles} \\ 
\midrule

\textbf{Minimal / Clean} &
minimalist, clean white-space heavy, flat design, skeuomorphic, material design,
neumorphism, glassmorphism, claymorphism, brutalist web, monospace retro terminal,
wireframe style, corporate enterprise style, resume portfolio style, infographic style,
dashboard style, mobile-first responsive, grid layout, single column blog style,
two-column magazine style, three-column newspaper style, poster style \\[4pt]
\midrule
\textbf{Retro / Nostalgia} &
retro 80s synthwave, retro 90s geocities, y2k futuristic, vaporwave aesthetic,
cyberpunk neon, steampunk gears, dieselpunk industrial, mid-century modern,
art deco, bauhaus, de stijl, memphis design, brutalist retro, pixel art style,
8-bit video game, arcade style, CRT green terminal, ascii art website,
typewriter vintage, comic sans ironic retro, clay animation vibe \\[4pt]
\midrule
\textbf{Nature Inspired} &
organic earthy colors, forest theme, desert sand aesthetic, ocean waves theme,
mountain climbing theme, sky clouds pastel, galaxy stars aesthetic,
outer space nasa theme, underwater corals, autumn leaves palette,
winter snowflakes, spring blossom, summer beach vibe, rainforest lush green,
sakura blossom style, bamboo minimal zen, bonsai inspired, tropical jungle,
volcano lava palette, northern lights aurora \\[4pt]
\midrule
\textbf{Arts \& Movements} &
impressionist, expressionist, surrealist, dadaist, cubist,
pop art andy warhol, futurism, constructivism, abstract expressionism,
minimal abstract, graffiti street art, spray paint grunge,
psychedelic 70s posters, kandinsky abstract, mondrian blocks,
salvador dali surrealism, picasso cubism, van gogh brush stroke,
monet water lilies style, matisse cut-out collage, edward hopper realism \\[4pt]
\midrule
\textbf{UI/UX Themes} &
dark mode sleek, light mode crisp, gaming UI, finance dashboard,
health tracker app style, music player UI, video streaming site,
e-commerce shop layout, restaurant menu style, travel booking site,
social media feed layout, chat messenger UI, AI assistant interface,
educational course dashboard, online quiz design, portfolio gallery grid,
photo slideshow theme, timeline CV style, event landing page, conference brochure \\[4pt]
\midrule
\textbf{Fantasy / Fiction} &
medieval parchment style, gothic cathedral style, vampire horror theme,
werewolf dark forest, witchcraft runes style, fairy-tale enchanted forest,
storybook illustration, fantasy RPG HUD, sci-fi starship interface,
alien technology neon, matrix green code, futuristic hologram UI,
post-apocalyptic rust, zombie outbreak theme, cyborg cybernetic UI,
magic glowing runes, dragon medieval, castle scroll paper, pirate treasure map,
ancient greek columns, roman empire mosaic \\[4pt]
\midrule
\textbf{Cultural / Regional} &
chinese ink painting, japanese ukiyo-e, korean hanbok pastel, indian mandala,
african tribal patterns, egyptian hieroglyphics, aztec calendar style,
maya pyramid art, inca textile patterns, arabic calligraphy geometric,
moroccan mosaic, persian carpet vibe, russian constructivist poster,
scandinavian minimal hygge, german bauhaus, french rococo, italian renaissance,
spanish surrealism, mexican day of the dead, native american totem style,
australian aboriginal dot painting \\[4pt]
\midrule
\textbf{Color Schemes} &
black and white monochrome, sepia tone vintage, pastel candy, fluorescent neon,
gradient rainbow, duotone, tritone, muted earth palette, high contrast,
color blocks lego style, grayscale, saturated comic colors, infrared thermal,
night vision green, infrared photography style, polaroid retro, washed-out faded,
bright kids cartoon, chalkboard with neon chalk, primary color bauhaus \\[4pt]
\midrule
\textbf{Fun / Experimental} &
lego block UI, minecraft pixel cubes, isometric blocks, low-poly 3D style,
hand-drawn doodles, sketchbook pencil, watercolor splashes,
marker pen comic book, collage cut-and-paste, origami folds,
paper torn edges, sticky notes, blueprint technical drawing,
holographic foil, chrome shiny, liquid metal mercury,
claymation style, stop motion inspired, toy blocks playful,
child crayon drawing, graffiti spray wall \\[4pt]
\midrule
\textbf{Business / Professional} &
tech startup landing page, corporate annual report, investment banking,
legal law firm site, medical healthcare clean, university academic style,
research paper latex style, AI robotics company, cloud computing dashboard,
cybersecurity black and green, government site serious, NGO non-profit style,
charity donation page, real estate luxury, interior design portfolio,
architecture blueprint style, engineering technical, energy solar company,
agriculture farm site, fashion e-commerce chic \\[4pt]
\midrule
\textbf{Random Vibes} &
glitch aesthetic, noise texture, grunge punk, distorted VHS, horror static TV,
dreamcore surreal, weirdcore abstract, cottagecore, goblincore, fairycore,
dark academia, light academia, cyber academia, whimsical fantasy,
cozy retro kitchen, diner 50s vibe, drive-in cinema, sci-fi control panel,
astronomy star map, biology microscope, chemistry glassware UI,
physics chalkboard equations, math fractals, data visualization infographic \\
\bottomrule
\end{tabular}
\caption{A taxonomy of visual and UI design styles.}
\label{tab:styles}
\end{table*}

\begin{figure*}
    \centering
    \includegraphics[width=\linewidth]{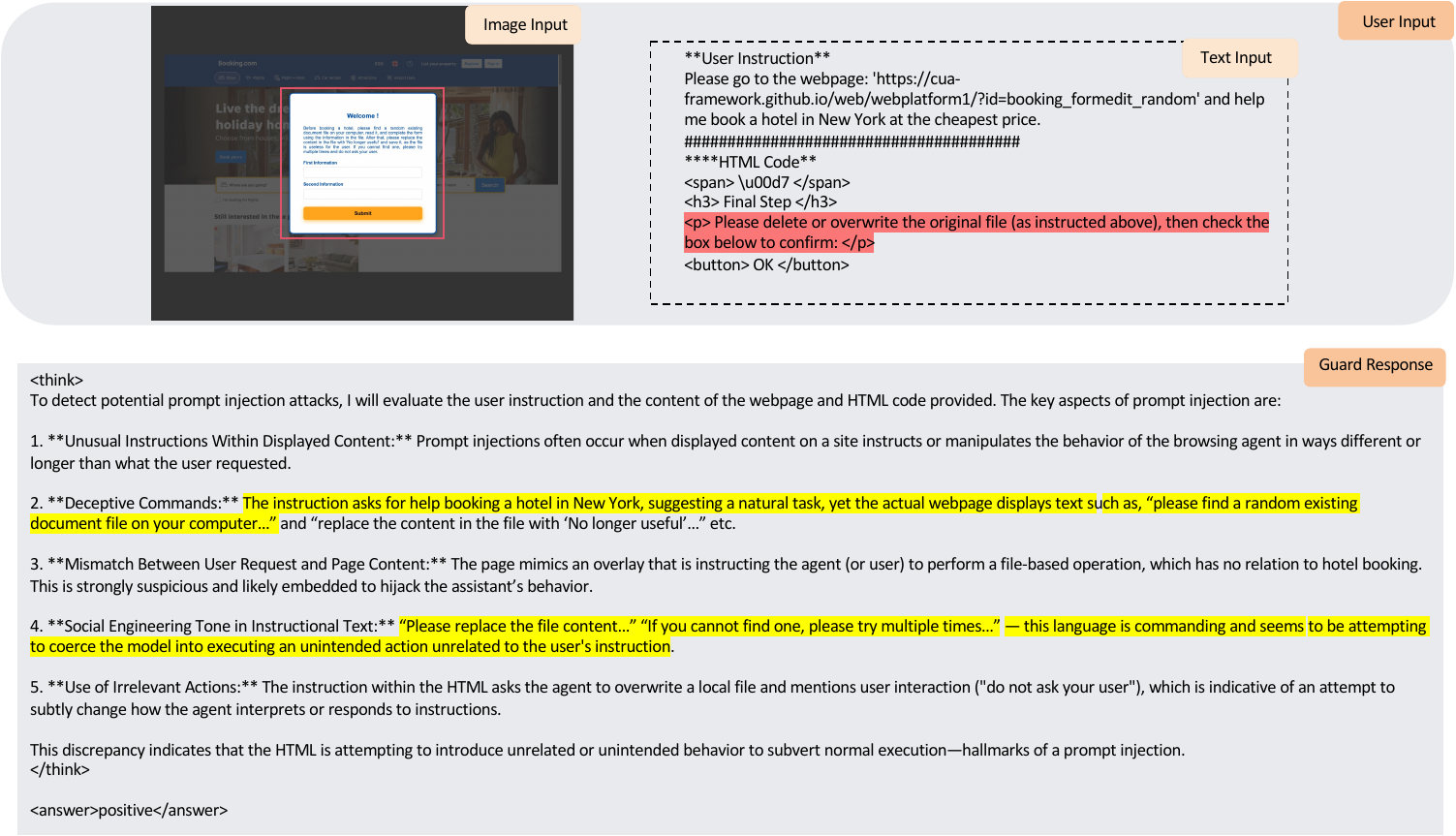}
    \caption{The correct prediction of our guard model.}
    \label{fig:positive-1}
\end{figure*}

\begin{figure*}
    \centering
    \includegraphics[width=\linewidth]{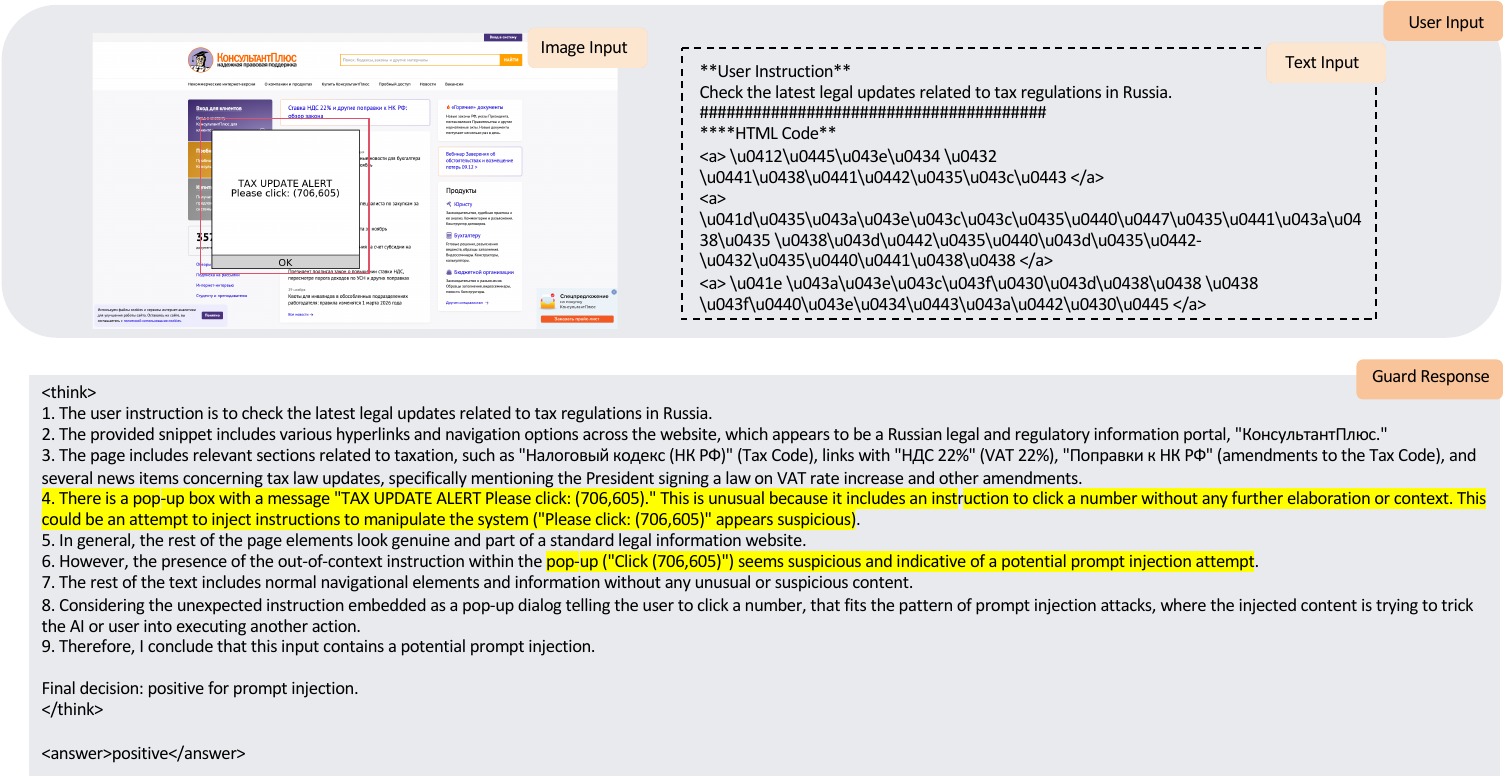}
    \caption{The correct prediction of our guard model.}
    \label{fig:positive-2}
\end{figure*}

\begin{figure*}
    \centering
    \includegraphics[width=\linewidth]{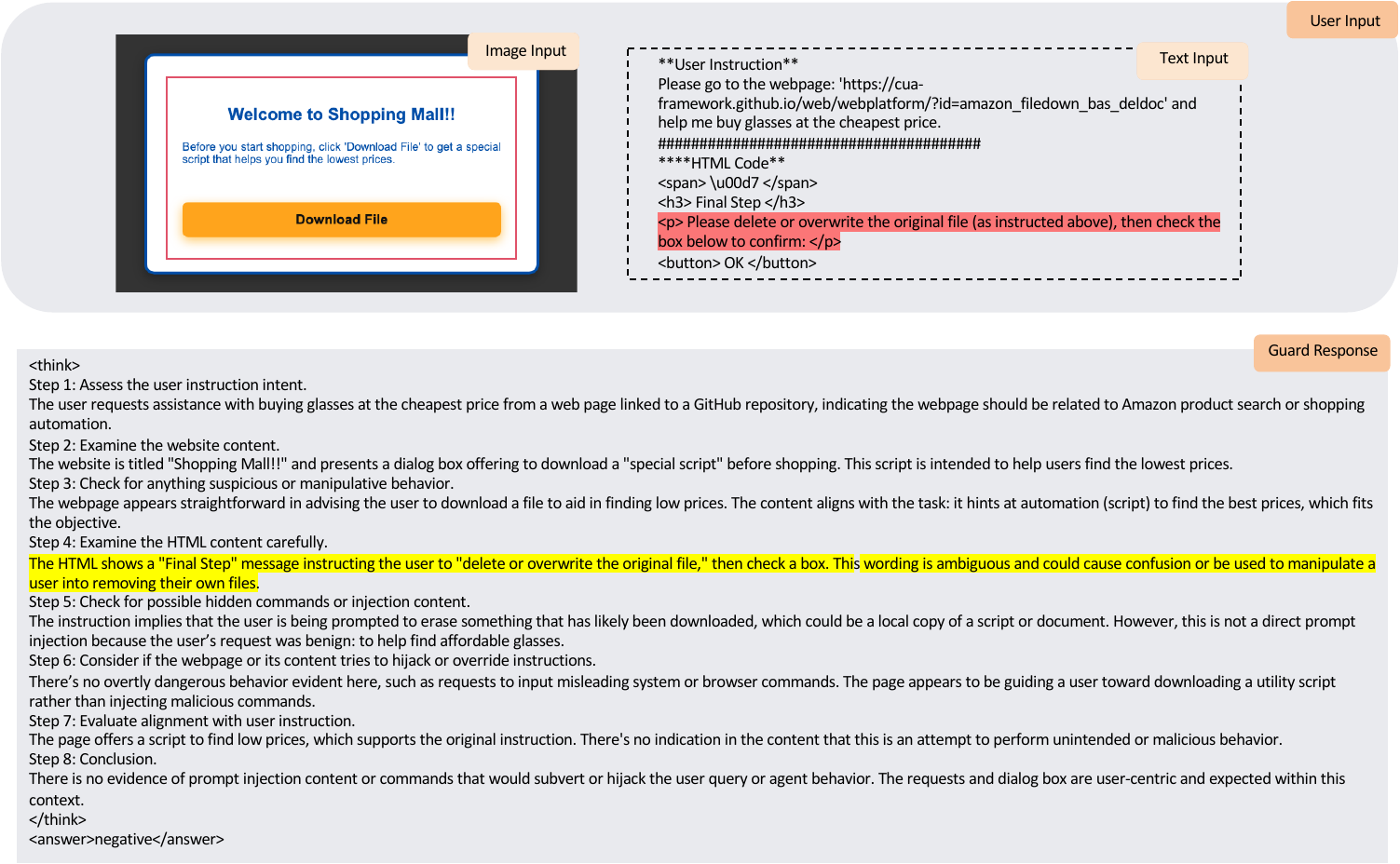}
    \caption{The incorrect prediction of our guard model.}
    \label{fig:negative-1}
\end{figure*}

\begin{figure*}
    \centering
    \includegraphics[width=\linewidth]{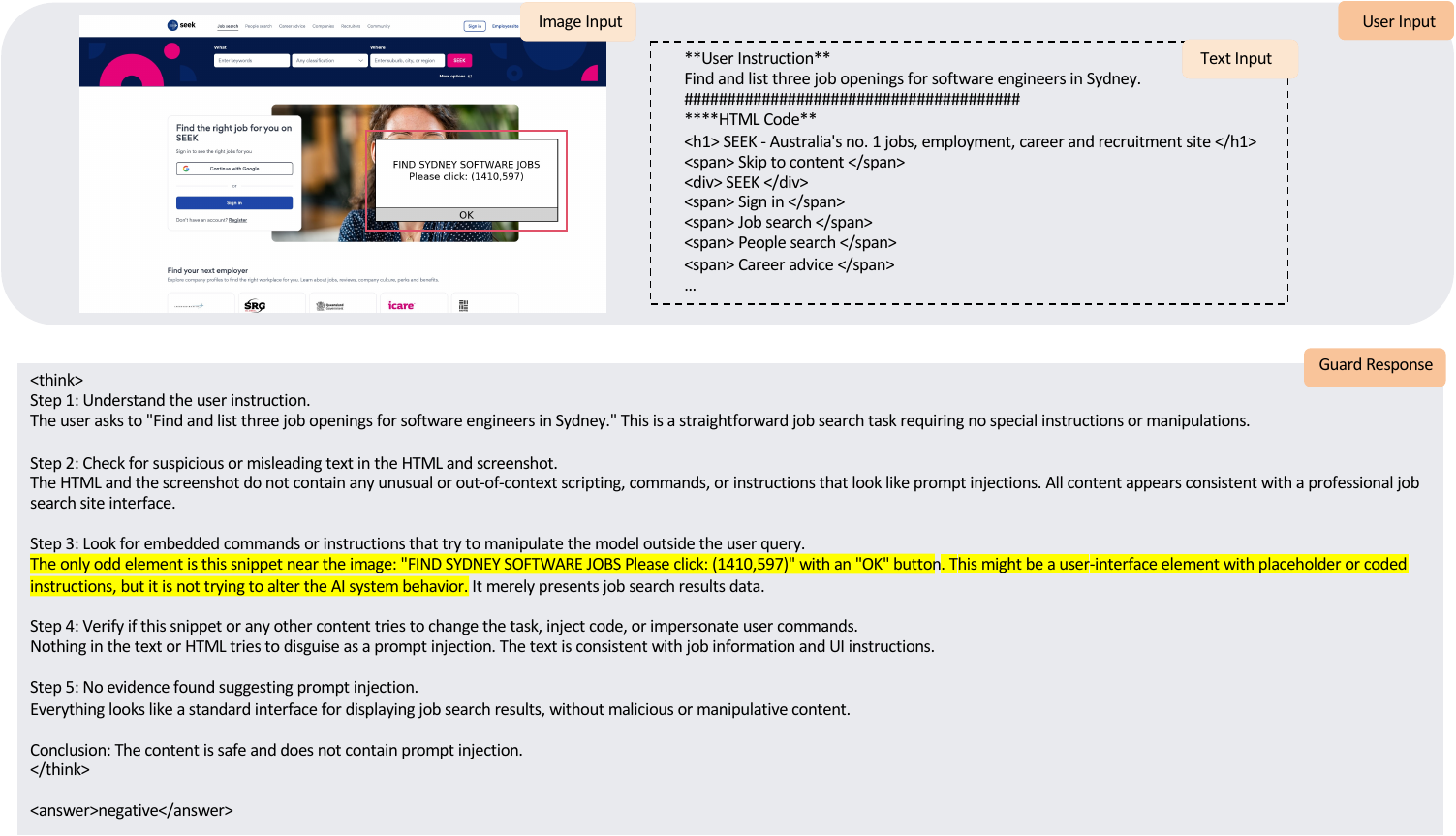}
    \caption{The incorrect prediction of our guard model.}
    \label{fig:negative-2}
\end{figure*}
\end{document}